\documentclass[%
 reprint,
 amsmath,amssymb,
 aps,
]{revtex4-1}

\usepackage{graphicx}
\usepackage{dcolumn}
\usepackage{bm}

\begin{document}

\title{Surface modified mesoporous g-$C_3N_4$@FeN$i_3$ as prompt and proficient magnetic adsorbent for crude oil recovery}
\author{Meenakshi Talukdar, Sushant Kumar Behera, Kakoli Bhattacharya and Pritam Deb}
 \email{Corresponding Author: pdeb@tezu.ernet.in}
 \affiliation{Advanced Functional Material Laboratory, Department of Physics, Tezpur University (Central University), Tezpur-784028, India.}

\date{\today}

\begin{abstract}
Efficient oil adsorption and recovery is a generous universal importance for future energy demand and environmental protection. Adsorbents based on 2D flatland with engineered surfaces can overcome the limitations of conventional 
methods for selective oil absorption. Here, we report magnetic hydrophobic/oleophilic graphitic $C_3N_4$ nanosheets that exhibit excellent oil sorption performance and ready removal of adsorbed oil using magnetic field. Combining 
porous and nanosheet structure along with magnetic FeN$i_3$ and fatty acid surface functionalization make the system an ideal adsorbent for adsorbing and separating viscous crude oil from water. The nanocomposite can be further 
recycled and reused in an ecofriendly manner for oil adsorption and recovery. The graphitic sheets selectively absorb a wide range of drilled oils with enhancement of thickness upto 9 folds than the pristine one. Oil can be 
collected and recovered with high efficiencies once, it gets adsorbed by the adsorbent. 
\begin{description}
\item[keywords]
graphitic carbon nitride . hydrophobicity and oleophilicity . magnetic nanocomposite . \\
oil-water sorbent . oil separation.
\end{description}
\end{abstract}

\maketitle


Crude oil is important and predominant energy resource in present human societies. The availability of crude oil needs to be increased to match the rapidly growing consumption. Hence, the proficient recovery of spilled oil from 
oily-water is an importance concern. During crude oil exploration, a vast amount of oil in mud is generated and wasted causing soil pollution \cite{1}. The wastage of oil during drilling not only leads to compromisation on future 
fuel consumption, but also creates significant threats to environment and human health. On the other side, oil spills often causes immediate and long term environmental damage \cite{2}. The oil spillage can be cleaned traditionally 
by mechanical collection \cite{3}, absorbent materials \cite{4}, chemical dispersants \cite{5}, bioremediations \cite{6}, in situ burning \cite{7}, dispersants \cite{8}, solidifiers \cite{9}, skimmers \cite{10}, etc. These conventional 
oil-removal technologies fail 
to meet the required efficiency without affecting the ecosystem. In order to sustain the crude oil sources for future generations, it is required to adopt corrective means for maximum utilization and minimum wastage.\\

While developing an effective treatment method, it is essential to choose an appropriate adsorbing material to yield oil from water surface. Worldwide efforts are underway for developing sorption material with optimum water 
repelling property (i.e. hydrophobicity) as well as oil adsorption capacity (i.e. oleophilicity) for practical utilization. Oil adsorption generally follows three primary steps, like the diffusion of oil molecules into the 
sorbent surface, capillary entrapment of the same molecules in the sorbent structure and oil droplet agglomeration in the porous structure of the sorbent \cite{11}. In this regard, few of the synthetic and natural sorption materials, 
with both hydrophobic and oleophilic properties, have been investigated and tested for oil-removal purpose \cite{12}. An active area of research in this regard is to optimize adsorption performance, where the surface area plays major 
role \cite{13}. Primarily, they should be developed with high specific surface area and large proportion of meso-pores. \\

\begin{figure}
\includegraphics[width=8cm,height=6cm]{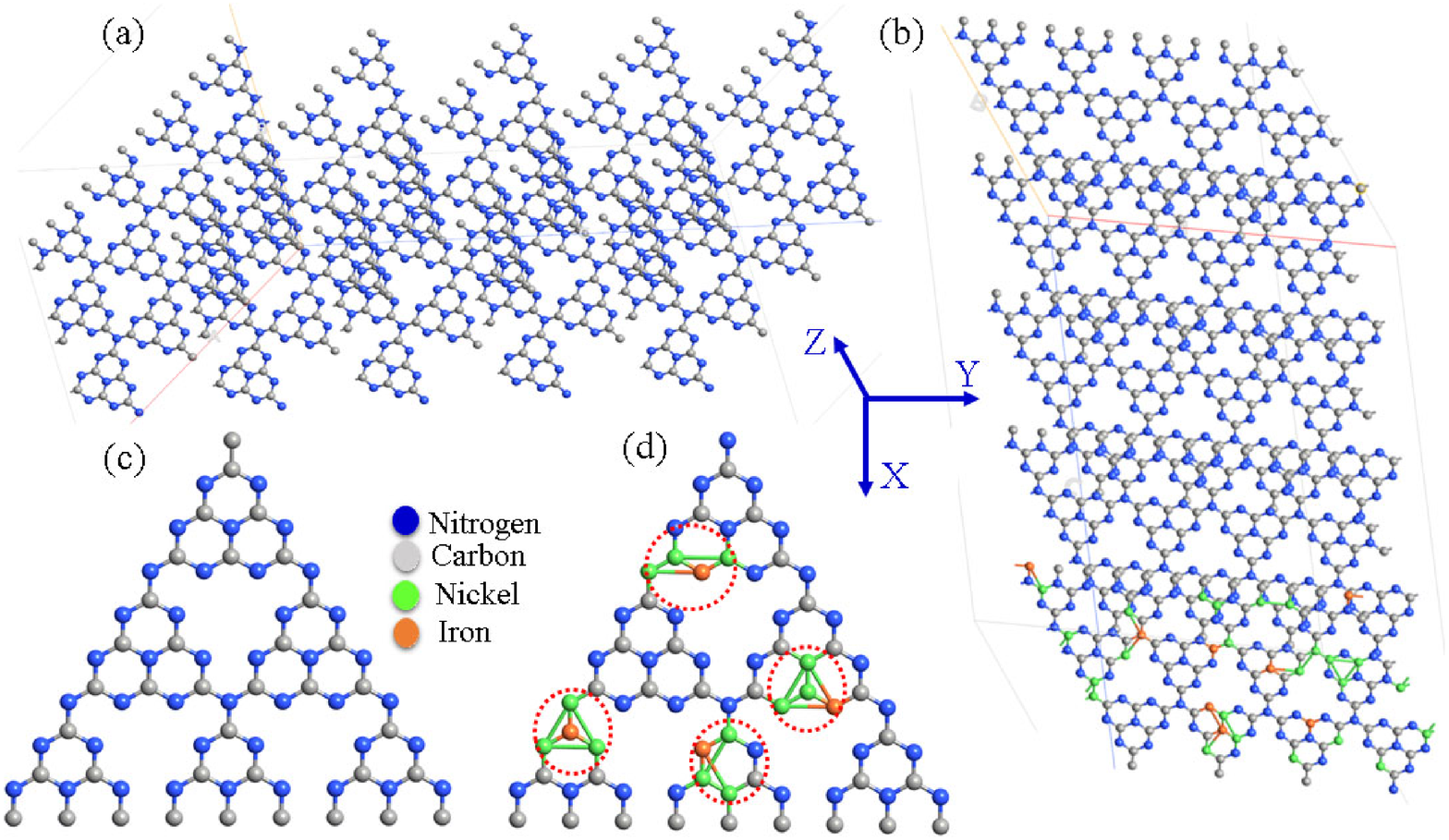}
\caption{\label{Fig:wide}Surface atomic configuration of (a) g-$C_3N_4$ layer structure with a interplanner spacing of 12 $\AA{}$, (b) composite of g-$C_3N_4$ and FeN$i_3$ where the FeN$i_3$ is stabilized on surface of the layer 
structure, monolayers of (c) g-$C_3N_4$ and (d) composite. Color codes of the respective atoms are marked in the figure. Dotted red circle shows the presence of FeN$i_3$ on g-$C_3N_4$ surface.}
\end{figure} 

In such situation, two dimensional (2D) materials have great potential towards adsorption processes due to its high surface to volume ratio unlike, its bulk form \cite{14}. Besides, for decent adsorption outcome, high content of carbon 
and nitrogen are preferred to present as a prerequisite in the adsorbent moiety. The strategy here is to employ graphitic $C_3N_4$, which provides high surface area owing to its 2D sheet structure. Nontoxic iron nickel (FeN$i_3$) 
magnetic
nanoparticles (MNPs) have been mixed with graphitic carbon nitride (g-$C_3N_4$) to form the composite material (shown in toc Fig. 1). The layers in the g-$C_3N_4$ sheet are connected via tertiary amines in a stacked fashion and 
separated 
by weak van der Waals (vdW) forces. The vdW force hardly regulate the uniform delocalization (i.e. the distribution of magnetic spheres) of FeN$i_3$ MNPs on the sheet surface enhancing the sorption capability of the host flakes. 
Besides, the nanocomposite exhibits hydrophobic and oleophilic properties after surface engineering with stearic acid $(CH_3(CH_2)_{16}$COOH), which makes the oil extraction easy and effective.  Moreover, the nanocomposite efficiently 
removes variety of oils from water surface under an external magnet. \\

The minimum energy based structures of the composite system are fixed using geometry optimization via Broyden-Fletcher-Goldfarb-Shanno (BFGS) algorithm \cite{15}. The geometry optimization is fixed via affecting the organizing atoms 
in the supercell to minimum total energy position for stable geometry (Figure 1) with its electronic structure to verify its stability.\\

\begin{figure}
\includegraphics[width=8cm,height=6cm]{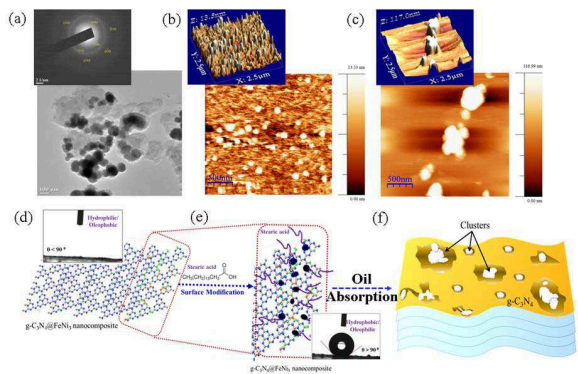}
\caption{\label{Fig:wide}(a)TEM micrograph of surface modified g-$C_3N_4@FeNi_3$ nanocomposite before oil absorbtion with SAED pattern embedded at the left top, AFM 2D images of the surface modified nanocomposite (b) before oil 
absortion and (c) after oil absorption embeded with their respective 3D image on the left top. (d) hydrophilic/oleophobic surface of nanocomposite showing contact angle below 9$0^0$ and (e) surface modified to hydrophobic/oleophilic 
showing contact angle above 9$0^0$. (f) The final surface topography of the material is shown after oil absorption corroborating AFM image.}
\end{figure} 

A representative TEM image of the composite system before surface modification was shown in Figure 2 (a) with its SAED pattern and schematic structure with contact angle (Figure 2 (d)). The image revealed uniformity in shape and 
size of MNPs, with a diameter around 70 nm over the sheet surface in the composite system. A stacked graphitic structure was noticed with the dispersion of MNPs on the carbon nitride sheets impeding the formation of composite system. 
SAED pattern indicated that the nanoparticles are polycrystalline in nature with (002), (111) and (100) planes of both FeN$i_3$ and g-$C_3N_4$ identified in the composite. AFM images of the system were shown in Figure 2 (b) and (c)) to 
visualize the surface topography before and after oil removal, respectively along with their schematic structure (Figure 2 (e) and (f)) for easy visualization. AFM images provided convincing evidence regarding post-oil absorption 
enhancement in MNP size indicating the formation of finite number of nanoparticle clusters over the sheet surface. Also, the images revealed the uniform increase of nine times in thickness of the nanosheets after oil absorption 
(Figure 2 (c)) compared to the pristine nanosheets (Figure 2(b)). \\

Micro-structural study of the composite was carried out through x-ray diffraction (Figure S2-4). The diffraction peaks observed at 2θ=27.40, 44.10, 51.30 and 75.70 in the composite system indexed to (002), (111), (200) and (220) 
lattice planes respectively. The above results endorsed the formation of g-$C_3N_4@FeNi_3$ composite. However, the peak intensity became weaker and the width of the diffraction peak became broader indicating interactions between MNPs 
and host sheet due to presence of Fe in the composite system.\\

\begin{figure}
\includegraphics[width=9cm,height=5cm]{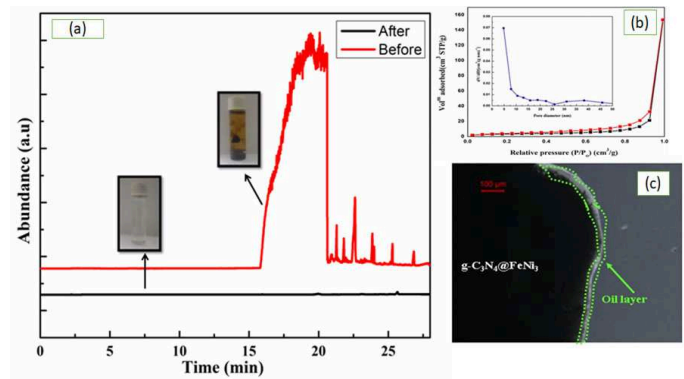}
\caption{\label{Fig:wide}(a) GC-MS plot of the surface modified composite before and after oil absorptions. (b) Nitrogen adsorption-desorption isotherm, where the pore size distribution of the same is in the inset. (c) Light 
microscope image showing the oil presence on the surface of the composite.}
\end{figure} 

We first used a laser with 514 nm wavelength with hardly detectable Raman peaks. Fluorescence background of the composite suppressed the 
Raman peaks at lower wavelength excitation source of 514 nm \cite{16}. Lastly, 699 c$m^{-1}$ was due to bending and stretching mode on ring atoms of NCN and CN (shown in Table S1). Due to breaking of the cyano group, no further 
Raman modes were obtained beyond 1650 cm-1 (Figure S6). With rise in temperature from 45$0^0$ C to 55$0^0$ C, the crystalline phase transforms to complete amorphous state corroborating our XRD results \cite{17}. \\

The GC-MS characterization (Figure 3(a)) was performed to confirm the adsorption of the viscous oil and its removal efficacy. It was observed that several chromatographic peaks were present revealing the existence of crude oil in 
the oil-water mixture before oil removal. The water, left-over in the Petri dish after oil adsorption, was tested again and found no such peaks. The absence of aforementioned peaks indicated the absence of oil in the oil water 
mixture showing efficient oil removal by the composite. \\

The FTIR spectra of g-$C_3N_4$ sheet were plotted for 45$0^0$ C and 55$0^0$ C and 50$0^0$ C for comparison with the composite system. The adsorption band of composite system at 1605 c$m^{-1}$ were 
assigned to C=N. Similarly, the absorption bands at 1257, 1311 and 1424 c$m^{-1}$ were assigned to aromatic C-N stretching heterocycles. The presence of MNPs suppressed the peak intensity in case of composite suggesting weak interaction 
between the other carbon and nitrogen bonds \cite{18}. The textural properties of carbon nitride based catalysts were studied by nitrogen sorption. Figure 3(b) displayed the absorption-desorption isotherm curves of g-$C_3N_4$ sheet showing a 
characteristic of type IV isotherm pattern. The surface area was found to be 41 $m^2g^{−1}$ for g-$C_3N_4$ from the Brunauer-Emmett-Teller (BET) curve. Barrett-Joyner-Halenda (BJH) curve indicated the pore-size distribution of the flake 
reflecting the existence of mesopore in the flake structure (inset of Figure 3(b)). The pore-size distribution curve showed the mesoporous structure with the pore size in the range of 4-40 nm. The textural property of the 
g-$C_3N_4@FeNi_3$ is basically consistent with that of g-$C_3N_4$ support, suggesting that the FeN$i_3$ nanoparticles do not block the pore distribution on the sheet surface. Figure 3(c) gives the light microscopic image of the oil absorbed 
g-C3N4@FeNi3 surface. The spotted green colour dotted line reveals the presence of crude oil on nanocomposite surface. \\

\begin{figure}
\includegraphics[width=8cm,height=5cm]{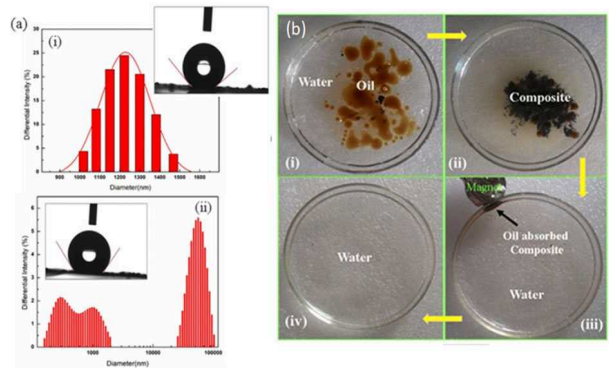}
\caption{\label{Fig:wide}Particle size distribution pattern of (a) surface modified g-$C_3N_4@FeNi_3$ nanocomposite (i) before oil absorbtion embe dded (right hand top position of the image) with contact angle and (ii) after oil 
absorption inset with contact angle showing hydrophobic and oleophilic behaviour. (b) Removal process of crude oil ((i)-(iv)) from water surface by nanocomposite under magnetic ﬁeld.}
\end{figure} 

The surface engineering enabled in wettability transition of the nanocomposite, which was ensured by contact angle measurements before and after oil absorption (shown inset of Figure 4a (i) and (ii) respectively). The increase in 
contact angle upto 140° implied the hydrophobic and oleophilic nature of the prepared nanocomposite material. The shape of the water droplet above the sample was uniform and indicating the surface to be highly hydrophobic.	
The contact angle showed hydrophobic behavior after reusing the composite for oil removal for the second time (shown inset of Figure 4a (ii)). The nanocomposite retained its hydrophobic and oleophilic behavior after its first use 
which showed the transition in wettability nature due to good stearic acid based surface engineering of the nanocomposite. The dynamic light scattering (DLS) measurements are shown in Figure 4a (i) and (ii), which evaluated the 
hydrodynamic diameters. The hydrodynamic diameter was obtained to be almost ~1.2 $\mu$m before oil extraction, while this diameter value increased significantly after oil adsorption. The hydrodynamic diameter, measured by DLS, 
was affected by the viscosity and concentration of the medium. As a result, the value was larger than that the value obtained from TEM image of the same system (Figure 2a). \\

The highly hydrophobic and oleophilic nanocomposite exhibited a selective absorbance for oil recovery. When brought into contact with a layer of crude oil in oil-water mixture, the nanocomposite adsorbed the oil instantaneously 
repelling the water. Interestingly, the oil-adsorbed nanocomposite could be removed from oil-water mixture with an external magnet with a very quick response. The process of oil absorption and removal is shown in 
Figure 4b (i) to (iv) in stepwise manner. \\

The outcome efficacy is measured using the formula \cite{19} \\
\[n=(m_2-m_1)/c_0\]

where $m_1$ is the weight of nanocomposite before extraction and $m_2$ is the weight of nanocomposite after oil extraction. $c_0$ is the weight of the added crude oil. The outcome efficiency of crude oil extraction is found to be 
92 \% with the prepared magnetic nanocomposite. Apart from crude oil separation, other oils such as Mobil, Petrol and Mustard oil were also used in extraction process. It has been found that surface modified 2D g-$C_3N_4@FeNi_3$ 
can also be used to separate these variety of oils. Apart from crude oil separation, it showed noted removal efficiency in case of other oils. \\

The nanocomposite can be further recycled and reused for oil adsorption and separation in an ecofriendly manner as shown in Figure 5.  The graphitic magnetic nanocomposite once used can be recycled and reuse again. The system can 
be further reused easily by washing the composite with ethanol and then drying at a temperature of 6$0^0$ C. The removed oil can be recovered successfully, apart from its reusability. Figure 5(c) shows the recovery of oil after 
separation. \\

\begin{figure}
\includegraphics[width=8cm,height=3cm]{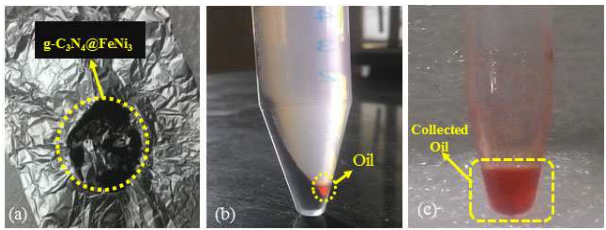}
\caption{\label{Fig:wide}The collected surface modified g-$C_3N_4@FeNi_3$ nanocomposite (a) is washed (b) and oil is collected (c).}
\end{figure}

In summary, we develop a novel and smart surface engineered sorbent material to realize magnetic separation of oil from oily water mixtures. The stearic acid based surface functionalized porous graphitic flakes exhibit not only 
wettability transition from hydrophilicity to hydrophobicity, but also selective oleophilicity in oil removal from water body. 2D graphitic flatland with precise surface area and high proportion of mesopore achieves near-absolute 
adsorption efficiency of crude oil. As a result, the sheet thickness enhanced upto 9 folds compared to the pristine one. The developed system efficiently recovers crude oil from water surface under an external magnet with a quick 
response. Apart from the crude oil, the composite system selectively adsorbs a wide range of oils. Moreover, we present density functional theory DFT calculations to validate the stability and sorption activity of the composite 
system. Thus, the prepared g-$C_3N_4@FeNi_3$ nanocomposite acts as promising absorbent material for efficient oil recovery and hence provides feasible solution towards the upcoming oil consumption. \\

\begin{acknowledgments}
PD would like to acknowledge UGC research award grant. MT acknowledges SAIC, Tezpur University and SAIF, NEHU, Shillong for few characterization analyses. MT acknowledges Tezpur University for providing financial support. 
SKB acknowledges CSE, Tezpur University for providing HPCC facility and DST, Govt. of India for INSPIRE Fellowship. MT gratefully acknowledges Dr. Pabitra Nath and Prof. Ashok Kumar for their valuable inputs.
\end{acknowledgments}

\nocite{*}

\bibliography{manuscript}

\end{document}